# Glucose availability but not changes in pancreatic hormones sensitizes hepatic AMPK activity during nutritional transition in rodents

Camille Huet[1], Nadia Boudaba[1], Bruno Guigas[2], Benoit Viollet[1], and Marc Foretz[1*]

From the [1]Université de Paris, Institut Cochin, CNRS, INSERM, F-75014 Paris, France and [2]Department of Parasitology, Leiden University Medical Center, Leiden, Netherlands.

Running title: *Hepatic AMPK regulation during nutritional transition*

*To whom correspondence should be addressed: Marc Foretz, Institut Cochin, Département d'Endocrinologie Métabolisme et Diabète, 24, rue du Faubourg Saint-Jacques, 75014 Paris, France. Phone: 33.1.44.41.24.38; Fax: 33.1.44.41.24.21; email: marc.foretz@inserm.fr

## Abstract

The cellular energy sensor AMP-activated protein kinase (AMPK) is a metabolic regulator that mediates adaptation to nutritional variations in order to maintain a proper energy balance in cells. We show here that suckling-weaning and fasting-refeeding transitions in rodents are associated with changes in AMPK activation and the cellular energy state in the liver. These nutritional transitions were characterized by a metabolic switch from lipid to glucose utilization, orchestrated by modifications in glucose levels and the glucagon:insulin ratio in the bloodstream. We therefore investigated the respective roles of glucose and pancreatic hormones on AMPK activation in mouse primary hepatocytes. We found that glucose starvation transiently activates AMPK, whereas changes in glucagon and insulin levels had no impact on AMPK. Challenge of hepatocytes with metformin-induced metabolic stress strengthened both AMPK activation and cellular energy depletion limited-glucose conditions, whereas neither glucagon nor insulin altered AMPK activation. Although both insulin and glucagon induced AMPKα phosphorylation at its Ser-485/491 residue, they did not affect its activity. Finally, the decrease in cellular ATP levels in response to an energy stress was additionally exacerbated under fasting conditions and by AMPK deficiency in hepatocytes, revealing metabolic inflexibility and emphasizing the importance of AMPK for maintaining hepatic energy charge. Our results suggest that nutritional changes (*i.e.* glucose availability), rather than the related hormonal changes (*i.e.* the glucagon:insulin ratio), sensitize AMPK activation to the energetic stress induced by the dietary transition during fasting. This effect is critical for preserving the cellular energy state in the liver.

## Introduction

AMP-activated protein kinase (AMPK) is a major energy sensor that regulates cellular and whole-body energy homeostasis (1). It is widely accepted that AMPK integrates nutritional and hormonal signals to maintain the cellular energy balance and execute appropriate metabolic functions (e.g., inhibition of ATP-consuming pathways and promotion of ATP-generating pathways) in response to nutritional environmental challenges. AMPK is activated in response to a variety of metabolic stresses or hormonal changes that typically change the cellular AMP/ATP and ADP/ATP ratios caused by increasing ATP consumption or reducing ATP production, such as that observed following starvation, exercise, hypoxia, ischemia, or inhibition of mitochondrial oxidative phosphorylation.

AMPK is a heterotrimeric complex consisting of a catalytic α subunit and two regulatory subunits, β and γ. Each subunit has at least two isoforms. The α-subunit contains the kinase domain, which is normally active only when a critical residue, Thr172, is phosphorylated within the activation loop (2). The upstream kinases that phosphorylate this site have been identified as the tumor suppressor liver kinase B1 (LKB1) and $Ca^{2+}$/calmodulin-activated protein kinase (CaMKK2). While Thr172 residue represents the major AMPK phosphorylation and activation site in the α-subunit, phosphorylation of some Ser/Thr residues within the ST loop by PKA, Akt, and GSK3, associated with reduced α-Thr172 phosphorylation, has been reported to inhibit AMPK activity (1,3). The β-subunit acts as a scaffold to link the three subunits and contains a myristoylation site that is important for the subcellular localization and activation of AMPK (4-6). The γ-subunit contains four tandem repeats of the cystathionine β-synthase (CBS) motif, which provides binding sites for the regulatory nucleotides, AMP, ADP, and ATP.

Binding of AMP or ADP activates AMPK by various mechanisms that are all inhibited by ATP. They include the promotion of AMPK α-subunit Thr172 phosphorylation by the upstream kinase LKB1 and inhibition of α-Thr172 dephosphorylation by protein phosphatases. In addition, binding of AMP, but not ADP, causes allosteric activation of up to 10-fold. Activation of AMPK can also occur independently of AMP/ADP binding through α-Thr172 phosphorylation by CaMKK2 in response to increased intracellular $Ca^{2+}$ levels. An additional AMP/ADP-independent mechanism is engaged upon glucose removal by the formation of an axin/LKB1/AMPK complex at the surface of lysosomes, leading to the phosphorylation and activation of a compartmentalized pool of AMPK. The activation of distinct subcellular pools of AMPK may play an important role in the phosphorylation of specific downstream targets. Indeed, a recent study reported that the intensity of stress stimulation triggers differential AMPK activation in the lysosomal, cytosolic, and mitochondrial fractions to target specific metabolic pathways, depending on the metabolic status of the cell (7).

In the liver, AMPK plays a crucial role in the regulation of lipid partitioning between oxidative and biosynthetic pathways through the phosphorylation and inactivation of its well-established targets, acetyl-CoA carboxylase (ACC) 1/2 at Ser79/Ser212 residue and 3-hydroxy-3-methylglutaryl (HMG) coenzyme A (CoA) reductase (HMGCR) at Ser871 residue (8-12). The transition from the fasting to refed state is associated with modifications in hepatic lipid metabolism (i.e., increased fatty-acid synthesis and decreased fatty-acid oxidation) that appear to coincide with changes in the activation and phosphorylation of the AMPK α-subunit at Thr172 and ACC at Ser79 (13-16). This observation raises the possibility that the modulation of AMPK activity may contribute to the shift of lipid metabolism in the liver from catabolism to anabolism. However, the specific cues that mediate such changes in AMPK signaling are still poorly understood. The hepatic metabolic adaptations that occur during fasting/refeeding are primarily triggered by changes in the glucagon/insulin ratio. During fasting, plasma glucagon levels are high and plasma insulin and glucose levels are low. By contrast, refeeding increases plasma glucose and insulin concentrations. Hence, changes in glucose availability and/or the level of pancreatic hormones may directly modulate hepatic AMPK activity during this metabolic transition. AMPK activity during fasting and refeeding may thus be regulated by glucagon or insulin-stimulated changes in kinase phosphorylation, respectively (15,17,18). Consistent with this possibility, the AMPK α-subunit is phosphorylated at multiple sites, including α1-Ser485/α2-Ser491, by the insulin-activated protein kinase Akt, inhibiting subsequent phosphorylation of α-Thr172 by upstream kinases (19,20).

Here, we provide evidence that hepatic AMPK activity is insensitive to changes in insulin and glucagon levels but is instead sensitive to variations in glucose availability. Such regulation is central to defining the threshold of AMPK activation during metabolic/energy stress in the liver.

**Results**

***Nutritional transition is associated with changes in AMPK activation and the energy state in the liver***

The suckling-weaning transition is accompanied by marked changes in metabolic pathways in the liver (i.e., a metabolic switch from lipid to glucose utilization with a decrease in lipid oxidation and an increase in glycolysis and lipogenesis) (21). During the suckling period, the plasma insulin concentration is low and that of glucagon high because of the ingestion of milk, which is a high-fat, low-carbohydrate food. The transition from a milk diet to a high-carbohydrate diet at weaning leads to an increase in blood glucose and plasma insulin levels (**Fig. 1A**). Weaning from maternal milk was associated with a decrease in the phosphorylation of AMPK at α-Thr172 and that of its substrate ACC at Ser79 in the liver, whereas the abundance of the AMPK and ACC proteins was unchanged (**Fig. 1B**). In this context, increased phosphorylation of Akt at Ser473 reflects activation of the insulin-signaling pathway (**Fig. 1B**). Conversely, AMPK signaling was highly active in the liver of suckling rats, as characterized by an increase in the phosphorylation of AMPK-α-Thr172 and ACC-Ser79 (**Fig. 1B**).

As expected, the transition from fasting to refeeding was associated with an increase in blood glucose and insulin levels and a decrease in blood glucagon levels (**Fig. 1C**). In the liver of starved mice, phosphorylation of AMPK-α-Thr172 and ACC-Ser79 was markedly higher than that of refed mice (**Fig. 1D**), in agreement with previous studies (13-16). Conversely, the increase in blood insulin levels in refed mice induced the phosphorylation of Akt at Ser473 residue (**Fig. 1D**). During fasting, changes in AMPK activation were associated with a lower cellular energy state, as revealed by the decrease and increase of hepatic ATP and ADP concentrations, respectively, resulting in a significant increase in the ADP/ATP ratio (**Fig. 2**). Thus, hepatic AMPK activation induced by fasting is associated with a decrease in the cellular energy state in the liver.

***AMPK deficiency exacerbates cellular energy depletion in response to metabolic stress in the liver***

AMPK plays a crucial role in maintaining energy homeostasis during periods of metabolic stress. We therefore hypothesized that AMPK deficiency in the liver may alter sensitivity to an energy stress. We tested this hypothesis by treating hepatic AMPK-deficient mice with metformin, a mitochondrial respiratory chain inhibitor (22,23). Metformin treatment induced a marked increase in the ADP/ATP ratio in the livers of control animals. Importantly, the increase in the ADP/ATP ratio was greater in AMPK-deficient liver (**Fig. 3A and 3B**). In agreement with the inhibitory action of metformin on mitochondrial complex 1 activity (22,23), we found that metformin treatment led to a reduction in respiration in control hepatocytes, which was accentuated in AMPK-deficient hepatocytes (**Fig. 3C**). These results demonstrate the protective role of AMPK in maintaining hepatic energy homeostasis in response to a metabolic challenge induced by a reduction in cellular energy charge.

***Metabolic stress-induced AMPK activation is strengthened in hepatocytes incubated under simulated fasting conditions***

Given the modulation of AMPK activity during the fasting/refeeding and suckling/weaning transitions, we hypothesized that the regulation of hepatic AMPK is driven by changes in glucose availability and/or glucagon or insulin-stimulated changes in the kinase phosphorylation status (15,17,18). We thus treated mouse primary hepatocytes with the AMPK activators, metformin, AICAR, or A-769662, under various nutritional and hormonal conditions mimicking the fasting or fed states to identify the nature of the stimuli that modulate hepatic AMPK activity.

Under low glucose/basal conditions (5 mM glucose), metformin induced the phosphorylation of AMPK at α-Thr172 and that of its downstream targets ACC at Ser79 and Raptor at Ser792 in concentration-dependent manner (**Fig. 4**). Incubation of hepatocytes with high glucose concentrations (25 mM glucose) and insulin, to mimic feeding, increased the phosphorylation of Akt at Ser473 but did not alter metformin-induced AMPK phosphorylation relative to that of the basal condition (**Fig. 4**). In contrast, culturing the hepatocytes in a medium that simulated fasting conditions, which contained glucagon and in which glucose was replaced with lactate and pyruvate, robustly enhanced metformin-induced AMPK phosphorylation concomitantly with an increase in the PKA substrate phosphorylation pattern (**Fig. 4**).

***Metabolic stress-induced energy depletion is worsened in AMPK-deficient hepatocytes incubated under simulated fasting conditions***

We next assessed the effect of changes in glucose concentrations and pancreatic hormone levels in culture medium on metformin-induced energy depletion in both control and AMPK-deficient hepatocytes. In the absence of metformin, the cellular energy charge was similar between control and AMPK-deficient hepatocytes and was not altered after an 8-h incubation under conditions mimicking fasting (glucose-free medium containing lactate and pyruvate plus glucagon) or refeeding (25 mM glucose plus insulin), indicating that these conditions are insufficient to alter the cellular energy state (**Fig. 5A and 5B**). Metformin treatment strongly correlated with a marked decrease in ATP and energy charge in both control and AMPK-deficient hepatocytes. As observed in the liver (**Fig. 3A and 3B**), the decrease in ATP levels and energy charge in response to metformin was greater in AMPK-deficient hepatocytes than control hepatocytes (**Fig. 5A and 5B**). Metformin-induced energy depletion was also greater when hepatocytes were incubated under fasting-like than refeeding-like conditions, and this effect was greater in AMPK-deficient hepatocytes than control hepatocytes. Thus, the higher activation of AMPK in response to an energy stress in hepatocytes incubated under fasting-like conditions coincides with a greater cellular energy deficiency. Moreover, the aggravation of energy depletion observed in AMPK-deficient hepatocytes incubated under fasting conditions is consistent with the role of AMPK in regulating the cellular energy balance to restore cellular ATP levels to normal values.

***Glucose availability, but not pancreatic hormones, sensitizes AMPK activation during metabolic stress in hepatocytes***

To dissociate the impact of glucagon and insulin signaling from that of glucose availability on AMPK activity, we separately examined their respective effects on metformin-induced AMPK activation in hepatocytes. In the presence of 5 mM glucose, treatment with glucagon or insulin induced sustained phosphorylation of PKA-substrates or Akt-Ser473, respectively, but they had no effect on the pattern of change of AMPK phosphorylation at α-Thr172 or that of its downstream targets (ACC at Ser79 and Raptor at Ser792) induced by metformin (**Fig. 6A**). Conversely, the incubation of hepatocytes in glucose-free medium containing lactate and pyruvate led to greater metformin-induced AMPK phosphorylation than that in hepatocytes incubated with 5 or 25 mM glucose for 8 h (**Fig. 6B**). We obtained similar results with A-769662, a direct small molecule AMPK activator (**Fig. S1**). Furthermore, AMPK and ACC phosphorylation induced by the cell-permeable AMPK activator AICAR was unaltered by treatment with the pancreatic hormones glucagon or insulin (**Fig. 7A**). In contrast, a comparison of the action of AICAR at various glucose concentrations showed AMPK phosphorylation to be stimulated to a greater extent in glucose-free medium containing lactate and pyruvate plus glucagon than medium containing only 5 mM glucose or 25 mM glucose plus insulin (**Fig. 7B**). In summary, AMPK activation in hepatocytes is enhanced by the scarcity of glucose, whereas changes in insulin or glucagon concentrations do not affect its activity.

***Switching to a glucose-free medium containing lactate and pyruvate transiently activates AMPK in hepatocytes***

In the previous experiments (**Fig. 4, 6, 7 and S1**), we observed no changes in AMPK-α-Thr172 phosphorylation after the incubation of hepatocytes for 8 h with various glucose concentrations in the absence of activators (metformin, AICAR or A-769662), This observation is consistent with the lack of an effect on the cellular energy state (**Fig. 5**). We tested whether changes in glucose availability could activate AMPK at early time points. Incubation of hepatocytes with a glucose-free medium containing lactate and pyruvate transiently activated AMPK signaling within 2 h (**Fig. 8A**). Indeed, phosphorylation of α-Thr172-AMPK and its downstream targets ACC and Raptor were maximal at 2 h and nearly returned to basal levels after 8 h (**Fig. 8A**). In contrast, incubation in a medium containing 5 or 25 mM glucose did not modify AMPK signaling (**Fig. 8A**). Moreover, AMPK activation induced by the absence of glucose in the culture medium correlated with a low but significant decrease in intracellular ATP levels at 2 h, which was amplified in AMPK-deficient hepatocytes (**Fig. 8B**). These results indicate that the lack of glucose induces transient activation of AMPK, which acts to adapt hepatocyte metabolism and maintain cellular energy levels.

We next assessed AMPK activation in response to various activators at early time points in hepatocytes incubated with various levels of glucose. Activation of AMPK by metformin (**Fig. 9**) or AICAR (**Fig. S2A**) was enhanced when hepatocytes were incubated in glucose-free medium containing lactate and pyruvate or with low glucose concentrations (5 mM glucose). Similarly, incubation of hepatocytes with the direct small molecule activator A-769662 induced more pronounced ACC and Raptor phosphorylation in glucose-free medium containing lactate and pyruvate (**Fig. S2B**). Unexpectedly, A-769662-mediated AMPK phosphorylation was only induced in the absence of glucose (**Fig. S2B**).

*Phosphorylation of AMPKα on the Ser485/491 residue does not alter its activity in hepatocytes*

Insulin and agents that elevate cellular cAMP have been reported to inhibit AMPK activity through the phosphorylation of AMPKα at Ser485/491 by Akt and PKA, respectively (20,24). We found that both insulin and glucagon weakly induced AMPK phosphorylation at α-Ser485/491 in primary hepatocytes. In contrast, AICAR induced massive phosphorylation at this site, likely resulting from autophosphorylation, as previously described (24). However, AMPK activity was not attenuated by increased α-Ser485/491 phosphorylation, as shown by maintenance of the phosphorylation of its downstream target ACC (**Fig. 7A and 7B**), indicating the absence of an inhibitory effect of Ser485/491 phosphorylation on AMPK activation in primary hepatocytes.

**Discussion**

Over the past decade, the fuel-sensing enzyme AMPK has attracted much attention because of the associations drawn between the wide range of its metabolic downstream targets, including fatty-acid synthesis and oxidation, mitochondrial function, oxidative stress, inflammation, and autophagy and the alteration of these pathways by insulin resistance and metabolic syndrome-associated disorders (25). Although there is no clear evidence that polymorphisms in genes encoding AMPK subunits influence the occurrence of metabolic syndrome (26-28), a sustained decrease in AMPK activity has been found in the liver, skeletal muscle, and adipose tissue from obese or hyperglycemic rodents and humans in association with insulin resistance (29-33). Nevertheless, we have shown that liver-specific AMPK deficient mice display normal hepatic glucose and lipid homeostasis and are not prone to insulin resistance, suggesting that the decrease in AMPK activity associated with insulin resistance may be a consequence, rather than a cause, of changes in hepatic metabolism (10).

By contrast, we and others have shown that reversible and physiological variation of hepatic AMPK activity occurs during the fasting-refeeding and suckling-weaning transitions (13-16) (**Fig. 1**). These changes in AMPK activity may account for the shift in hepatic lipid metabolism from catabolism to anabolism. However, the catabolic and/or anabolic stimuli responsible for the physiological modulation of AMPK activity during fasting are still poorly understood. It has been hypothesized that acute changes in hepatic AMPK activity are due to fluctuations in plasma levels of insulin and the counter regulatory action of glucagon (34). Indeed, conditions associated with increased glucagon activate AMPK, possibly through modulation of the hepatic energy charge (increase in the AMP/ATP ratio) and PKA-induced activation of LKB1 (18,35). Conversely, insulin has been reported to decrease AMPK activity through the phosphorylation of AMPKα at α1-Ser485/α2-Ser491, with a concomitant loss of both AMPKα-Thr172 and ACC-Ser79 phosphorylation (17,36).

The role of the phosphorylation at Ser485/491 is not well understood and it is still unclear whether this phosphorylation contributes to enzyme regulation and AMPK activity. Interestingly, mutation of the Ser485 residue to mimic phosphorylation by introduction of an aspartate residue in the AMPKα1 subunit is not sufficient to inhibit AMPK activation by liver purified AMPK kinase (37). Although phosphorylation at Ser485/491 has been shown to correlate with the inhibition of AMPK activity in a variety of tissues, we show that increased phosphorylation at this site is proportional to the increase in the phosphorylation of the downstream AMPK target ACC-Ser79 in primary hepatocytes (**Fig. 7**). Similarly, HepG2 cells treated with troglitazone showed an increase in phosphorylation of AMPKα at both Thr172 and Ser485 residues, associated with an increase in ACC-Ser79 phosphorylation (38). This is also reminiscent of the effect of acute renal ischemia causing simultaneous phosphorylation of AMPKα-Thr172 and AMPKα-Ser485 in the kidney (39). Furthermore, despite that glucagon and insulin pretreatments induce phosphorylation at α-Ser485/491, this is not sufficient to reduce the phosphorylation of the activation loop α-Thr172 in response to AICAR (**Fig. 7**). It is likely that the increase in Ser485/491 phosphorylation reflects an autophosphorylation event, concomitant with the increase in AMPK activation, as previously demonstrated (24). It was suggested that phosphorylation of AMPK at this specific site may represent a regulatory mechanism to prevent over-stimulation of AMPK, in addition to potential cross-talk with inhibitory signaling pathways (24). Thus, further studies will be required to better understand the physiological impact of such phosphorylation on hepatic metabolism.

We and others have shown that AMPK activation in the liver during fasting results from an increase in AMP/ATP and ADP/ATP ratios (40,41) (**Fig. 2**). However, the nature of the stimuli altering the AMP/ATP and ADP/ATP ratios and subsequent AMPK activity in the liver is somewhat unclear. Our results suggest that nutritional changes (i.e., glucose availability) rather than related hormonal changes (i.e., glucagon/ insulin ratio) likely underly the sensitization of AMPK to energetic stress induced by the dietary transition that takes place during fasting.

Recent studies suggest that starvation-induced AMPKα-Thr172 phosphorylation in the liver requires the formation of a ternary complex between axin, LKB1, and AMPK (16). Interestingly, AMP binding to AMPK has been shown to enhance its binding to the axin-LKB1 complex and thus promote axin-dependent AMPKα-Thr172 phosphorylation (7,16). Also, it has been reported that AMPK can sense glucose starvation independently of changes in adenine nucleotide concentrations through the formation of a lysosomal complex (42). Although we were unable to detect activation of AMPK in primary mouse hepatocytes incubated in glucose-free medium containing lactate and pyruvate for 8 h, we observed transient AMPK activation within 2 h, associated with a decrease in cellular ATP levels (**Fig. 8**). Such transient activation of AMPK observed in hepatocytes after switching from glucose to lactate and pyruvate (**Fig. 8**) can be seen as a counter signal to adapt the metabolism to glucose starvation and maintain the cellular energy charge. Of note, transient AMPK activation induced by glucose starvation was low compared to drug-induced AMPK activation (**Fig. 9 and Fig. S2**), suggesting that substitution of glucose by lactate and pyruvate in medium had a relatively low impact on cellular energy levels.

Furthermore, consistent with its role as an energy sensor acting to restore energy homeostasis, primary hepatocytes incubated in the absence of glucose (but incubated with lactate and pyruvate) exhibited enhanced activation of AMPK and amplified energy depletion in response to metformin-induced energy stress relative to that of hepatocytes incubated with glucose (5 or 25 mM) (**Fig. 4, 5, 6 and 9**). Such an enhanced response to energy stress observed in the context of low glucose availability may be attributable to lower ATP generation due to an overall decrease in the glycolytic flux.

Unexpectedly, we showed that the stimulation of AMPK signaling by the small molecule A-769662 was enhanced by low glucose levels and that AMPKα-Thr172 phosphorylation was transiently induced under conditions of glucose starvation (**Fig. S1 and S2**). These effects appear to be paradoxical since A-769662 causes activation of AMPK independently of α-Thr172 phosphorylation via an allosteric mechanism and without compromising the cellular AMP/ATP ratio (43-45). Nevertheless, we have previously shown that AMP-induced phosphorylation of AMPK is enhanced by A-769662 (10,44). Thus, the AMPK-α-Thr172 phosphorylation observed with A-769662 in glucose-free medium can be interpreted as a synergic effect resulting from AMPK-α-Thr172 phosphorylation caused by a transient increase in cellular AMP levels in response to glucose starvation and A-769662-binding to the AMPK complex.

ATP depletion was more pronounced in the livers of hepatic AMPK-deficient mice treated with metformin than in those of control mice (**Fig. 3**). Similarly, ATP levels in primary AMPK-deficient hepatocytes were much lower than in control hepatocytes following incubation with metformin (**Fig. 5**), as previously reported (44). The depletion of hepatic ATP observed during fasting has also been shown to be amplified in the livers of AMPK-deficient mice (41). This exacerbated decrease in ATP levels in response to an energy stress reveals metabolic inflexibility in AMPK-deficient hepatocytes and emphasizes the importance of AMPK in the maintenance of the hepatic energy charge through the control of adaptive mitochondrial function (46-49). Of note, in the absence of energy stress, ATP levels in the livers and primary hepatocytes from AMPK-deficient mice are no different than those of the control counterparts, reinforcing the notion that hepatic AMPK is activated only during times of energy stress to maintain the energy balance. These results demonstrate that AMPK activation is crucial for maintaining energy homeostasis in the liver during the metabolic transition that occurs during fasting.

AMPK has a high therapeutic potential for the management of dysregulated metabolism in the liver. Notably, pharmacological AMPK activation has shown beneficial effects in the treatment of liver steatosis (10,43,50,51). We have shown that drug-induced AMPK activation decreases hepatic lipid accumulation, both by inhibiting lipid synthesis and by stimulating fatty-acid oxidation (10). Thus, the modulation of hepatic AMPK activity by the nutritional state may have implications in future clinical practice. Indeed, our findings predict that hepatic AMPK activation in response to the administration of an AMPK-activating drug may be enhanced during fasting. On the other hand, postprandial delivery of an AMPK-activating compound may counteract the lowing of hepatic AMPK activity due to the massive influx of glucose into liver after the ingestion of a carbohydrate-rich meal. In both conditions, the nutritional state may influence the downstream effects of AMPK. In the fasting state, AMPK-induced fatty-acid oxidation will be boosted. In the fed state, lipid synthesis from glucose will be inhibited by drug-induced AMPK activation. Consequently, the delivery of future AMPK activating therapies will need to consider the nutritional state and diet composition (low- or high-carbohydrate) to adapt the dosage.

In summary, our studies support the notion that reversible AMPK activation observed in the liver during nutritional transition (typically during fasting) results in a decrease in cellular energy charge, which is modulated by glucose availability rather changes in pancreatic hormone levels. In this context, AMPK activation is critical to preserve the cellular energy state in the liver by promoting a metabolic switch from the utilization of glucose to that of other substrates, notably, fatty acids, to supply energy needs.

**Experimental procedures**

*Reagents and antibodies*

Metformin (#D5035) and glucagon were purchased from Sigma. AICAR was purchased from Toronto Research Chemicals. A-769662 was kindly provided by Dr. Anudharan Balendran (Astra Zeneca). Human insulin (Actrapid) was purchased from Novo Nordisk. Primary antibodies directed against total AMPKα (#2532), AMPKα phosphorylated at Thr172 (#2531), phospho-AMPKα1(Ser485)/ AMPKα2(Ser491) (#4185), total acetyl-CoA carboxylase (ACC) (#3676), ACC phosphorylated at Ser79 (#3661), total Raptor (#2280), Raptor phosphorylated at Ser792 (#2083), total Akt (#9272), Akt phosphorylated at Ser473 (#4058), and phospho-PKA substrate (#9624) were all purchased from Cell Signaling Technology. HRP-conjugated secondary antibodies were purchased from Calbiochem. All other materials unless otherwise indicated were purchased from Sigma.

*Animals*

Animal studies were approved by the Paris Descartes University ethics committee (no. CEEA34.BV.157.12) and performed under French authorization to experiment on vertebrates (no.75-886) in accordance with the European guidelines. C57BL/6J mice were obtained from Harlan France. Liver-specific double knockout of AMPKα1 and AMPKα2 catalytic subunits was achieved by crossing AMPKα1$^{lox/lox}$ mice with AMPKα2$^{lox/lox}$ mice and then crossing the progeny with Alfp-Cre transgenic mice to generate AMPKα1$^{lox/lox}$, α2$^{lox/lox}$ (control) and AMPKα1$^{lox/lox}$, α2$^{lox/lox}$-Alfp-Cre (liver AMPKα1/α2 KO) mice (10). All mice were maintained in a barrier facility under a 12/12-h light/dark cycle with free access to water and standard mouse diet (in terms of energy: 65% carbohydrate, 11% fat, 24% protein).

*Suckling and weaned rats*

Litters of 13-day-old Wistar rats with their mother were obtained from Janvier France. Litters were housed in individual cages under a 12/12-h light/dark cycle in a temperature-controlled environment with free access to water and standard diet. When the pups were 18 and 19 days of age, the mothers were fed separately from their offspring from 9:00 to 12:00 a.m. and 4:00 to 7:00 p.m. to avoid early weaning of the pups. Twenty-day-old suckling pups were separated from their mother for 3 h to allow gastric emptying. Then, rats were force-weaned by gavage with 5 g/kg glucose or replaced with their mother. After 3 h, rats were sacrificed by decapitation. Blood was quickly collected, and the liver was immediately removed and frozen in liquid nitrogen in < 25 s. Livers were stored at -80°C until analysis. After sacrifice, the stomachs of suckling pups were checked to ensure that they were filled with milk.

*Fasting and refeeding experiments*

For the fasting-refeeding experiments, mice were fasted for 24 h or fasted for 24 h and then refed a high-carbohydrate diet (70% carbohydrate in terms of total kcal with 64% sucrose in terms of weight, Harlan TD.08247) for 3 h. At the end of the refeeding period, mice were sacrificed by cervical dislocation and the liver immediately removed and frozen in liquid nitrogen in < 25 s. Livers were stored at -80°C until analysis.

*Blood glucose and plasma pancreatic hormone measurement*

Blood glucose concentrations were determined from blood isolated from the tail vein with a glucometer (Roche Diagnostics). Blood was collected into heparin-containing tubes and centrifuged to obtain plasma. Plasma insulin and glucagon levels were determined using mouse or rat ELISA kits (Crystal Chem).

*In vivo metformin treatment*

Ten-week-old control and liver AMPKα1/α2 KO mice in a fed state were injected intraperitoneally with saline or 200 mg/kg metformin to induce a hepatic energy stress. Mice were sacrificed by cervical dislocation 1 h after metformin administration and the liver extracted and frozen in liquid nitrogen in < 25 s. Livers were stored at -80°C until adenine nucleotide analysis.

*Mouse primary hepatocytes*

Mouse primary hepatocytes were isolated from 10-12-week-old male mice using a modified version of the collagenase method as described previously (44). The cells were plated in M199 medium with Glutamax supplemented with 100 U/ml penicillin, 100 μg/ml streptomycin, 10% (v/v) FBS, 500 nM dexamethasone (Sigma), 100 nM triiodothyronine (Sigma), and 10 nM insulin (Actrapid, Novo Nordisk) at a density of 4 x $10^5$ cells/well in six-well plates or 1 x $10^6$ cells/60-mm-diameter cell culture plate. After attachment (3 to 4 h), hepatocytes were maintained in M199 medium with antibiotics and 100 nM dexamethasone for 16 h. The cells were then stimulated with the respective compounds or hormones for the times indicated in the figure legends in glucose-free DMEM medium supplemented with 100 nM dexamethasone and lactate/pyruvate (10:1 mM) or glucose (5 or 25 mM).

*Hepatocyte oxygen consumption assay*

Primary mouse hepatocytes were plated onto collagen I-coated Seahorse 96-well plates at a density of 10,000 cells/well. After 4 h, primary hepatocytes were cultured for 16 h in M199 medium containing antibiotics and 100 nM dexamethasone. Hepatocytes were then switched to glucose-free DMEM supplemented with lactate/pyruvate (10:1 mM) and 100 nM dexamethasone 1 h prior to measuring respiration. The oxygen consumption rate (OCR, mitochondrial respiration) was monitored using the Seahorse Bioscience XF96 Extra Cellular Flux Analyzer in real time. The OCR was acquired under basal conditions and 15, 30, and 45 min after injection with 1 mM metformin. Results were normalized to total protein/well after completion of the assay and are expressed as pmoles $O_2$ consumed per μg of protein per minute.

*Measurement of adenine nucleotide concentrations*

Adenine nucleotide concentrations were determined in cell extracts prepared from cultured hepatocytes or liver samples using an enzymatic method (44). Primary hepatocytes were treated as described in the figure legends, the culture medium removed, and cells on 60-mm-diameter cell culture plates (1 x $10^6$ cells/plate) scraped into 200 μl 6% (v/v) ice-cold $HClO_4$ in < 5 s. For the liver, mice were treated as described in the figure legends. At the end of treatment, mice were sacrificed by cervical dislocation and the livers extracted and frozen in liquid nitrogen in < 25 s. Two hundred milligrams of liver were homogenized in 1 ml 6% (v/v) ice-cold $HClO_4$. Cell extracts were centrifuged at 10,000 g for 10 min at 4°C. The acid supernatant was neutralized and used for spectrophotometric determination of adenine nucleotides. Standard curves for ATP, ADP, and AMP were constructed with 25, 50, 75, 100, 125, and 150 µM of each nucleotide. Determination of the adenine nucleotides presented in the **Fig. 8B** was performed by high-performance liquid chromatography as described previously (52). Adenine nucleotide levels are expressed in µmol/g of liver weight or nmol/mg of protein. The energy charge was calculated using the following equation: [ATP+ADP/2]/[ATP+ADP+AMP], where AMP, ADP, and ATP are the respective tissue concentrations (53).

*Western-blot analysis*

After the incubation times indicated in the figure legends, cultured hepatocytes were lysed in ice-cold lysis buffer containing 50 mM Tris, pH 7.4, 1% Triton X-100, 150 mM NaCl, 1 mM EDTA, 1 mM EGTA, 10% glycerol, 50 mM NaF, 5 mM sodium pyrophosphate, 1 mM $Na_3VO_4$, 25 mM sodium-β-glycerophosphate, 1 mM DTT, 0.5 mM PMSF, and protease inhibitors (Complete Protease Inhibitor Cocktail; Roche). Lysates were sonicated on ice for 15 seconds to shear DNA and reduce viscosity. Pieces of liver were homogenized in ice-cold lysis buffer with a ball-bearing homogenizer (Retsch). The lysates and homogenates were centrifuged for 10 min at 10,000 x g at 4°C and the supernatants removed for determination of total protein content with a BCA protein assay kit (Thermo Fisher Scientific). Fifty micrograms of protein from the supernatant was separated on 10% SDS-PAGE gels and transferred to nitrocellulose membranes. The membranes were blocked for 30 min at 37°C with Tris-buffered saline supplemented with 0.05% NP40 and 5% nonfat dry milk. Immunoblotting was performed with the antibodies indicated in the figure legends, following standard procedures, and the signals detected by chemiluminescence reagents (Thermo). Total and phosphorylated AMPK, ACC, Raptor, and Akt were probed using separate membranes. X-ray films were scanned and band intensities were quantified by Image J (NIH) densitometry analysis.

*Statistical analysis*

Results are expressed as the means ± SD. Comparisons between groups were made by unpaired two-tailed Student's t-tests or one-way ANOVA, in conjunction with Bonferroni's post hoc test for multiple comparisons, when appropriate, using GraphPad Prism 5.0 (GraphPad Software Inc.). Differences between groups were considered statistically significant when $P < 0.05$.

**Data Availability Statement:** All data presented and discussed are contained within the manuscript.

**The abbreviations used are:** AMPK, AMP-activated protein kinase; ACC, acetyl-CoA carboxylase; AICAR, 5-aminoimidazole-4-carboxamide ribonucleotide; L/P, glucose-free medium containing 10 mM lactate and 1 mM pyruvate; G5, medium containing 5 mM glucose; G25, medium containing 25 mM glucose.

**Financial support:** This work was supported by grants from Inserm, the CNRS, the Université Paris Descartes, the Région Ile-de-France (CORDDIM), and the Société Francophone du Diabète (SFD). N.B. is a recipient of a doctoral fellowship from the French Government (Ministère de la Recherche et des Enseignements Supérieurs).

**Author contributions:** CH and NB performed experiments. BG performed the adenine nucleotide measurement by HPLC and edited the manuscript. BV interpreted the data and wrote the manuscript. MF conceived, designed, and performed experiments, interpreted the data, wrote the manuscript, and directed this study.

**Conflict of interest:** The authors declare no conflict of interest associated with this study.

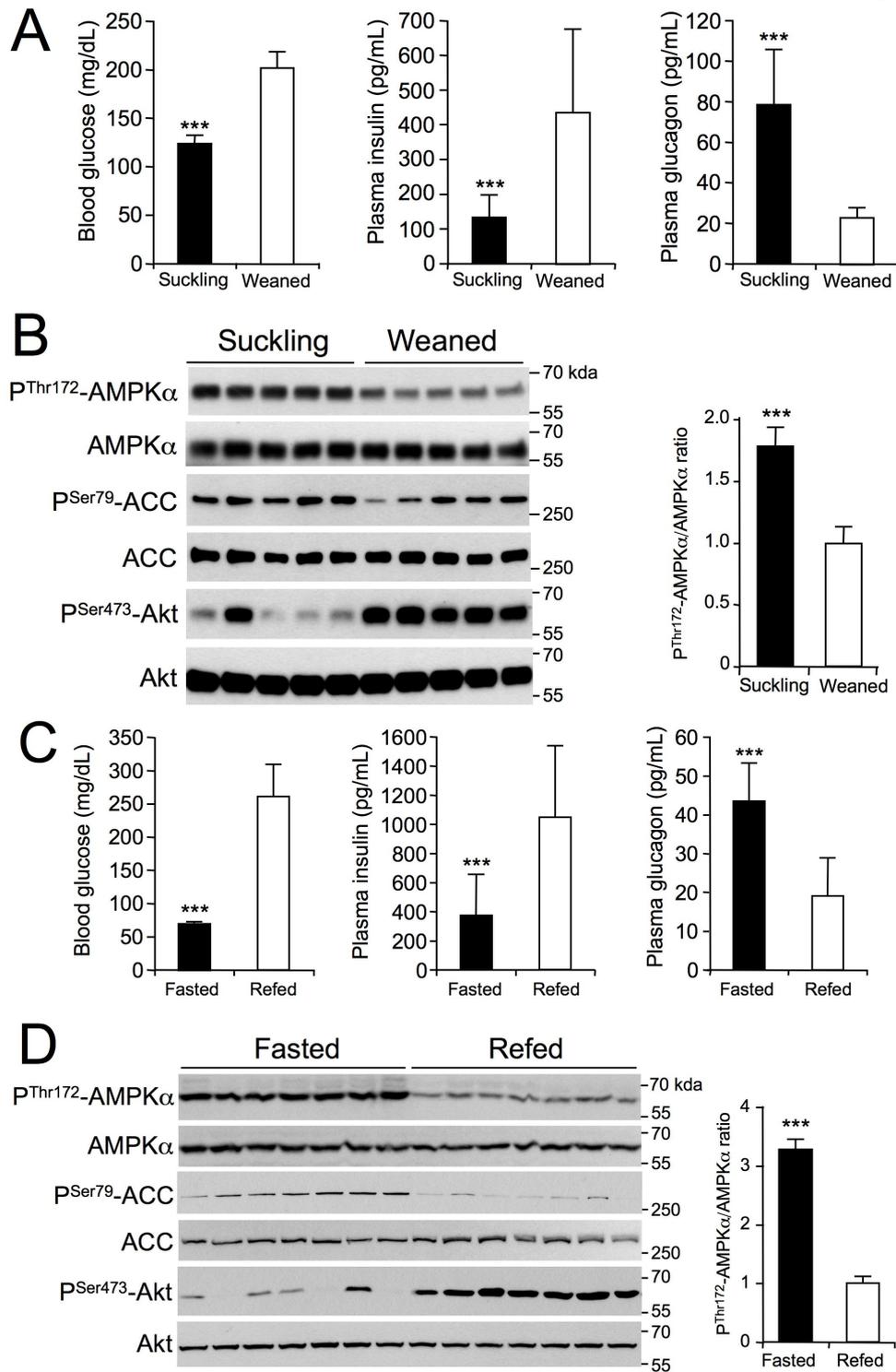

**Figure 1. Effect of suckling/weaning and fasting/refeeding transitions on AMPK activation in the liver.** (**A, B**) Twenty-day-old suckling rats were separated from the mother for 3 h. They were then either force-weaned by gavage with 5 g/kg glucose (Weaned) or placed back with the mother (Suckling) for 3 h. (**C, D**) Ten-week-old C57BL6J mice were either fasted for 24 h (Fasted) or fasted for 24 h and then refed a high-carbohydrate diet (Refed) for 3 h. After nutritional manipulation, (**A, C**) blood glucose levels were determined, blood was collected to assess plasma insulin and glucagon levels (n = 10-12 per group), and (**B, D**) the livers were quickly collected for western-blot analysis using the indicated antibodies. Each lane represents the liver sample from an individual animal. Right panels represent the P-Thr172-AMPKα/AMPKα ratio from the quantification of immunoblot images (n = 5-6 per group). Data are presented as the means ± SD. ***$P < 0.001$ compared to refed mice or weaned rats.

# Figure 2

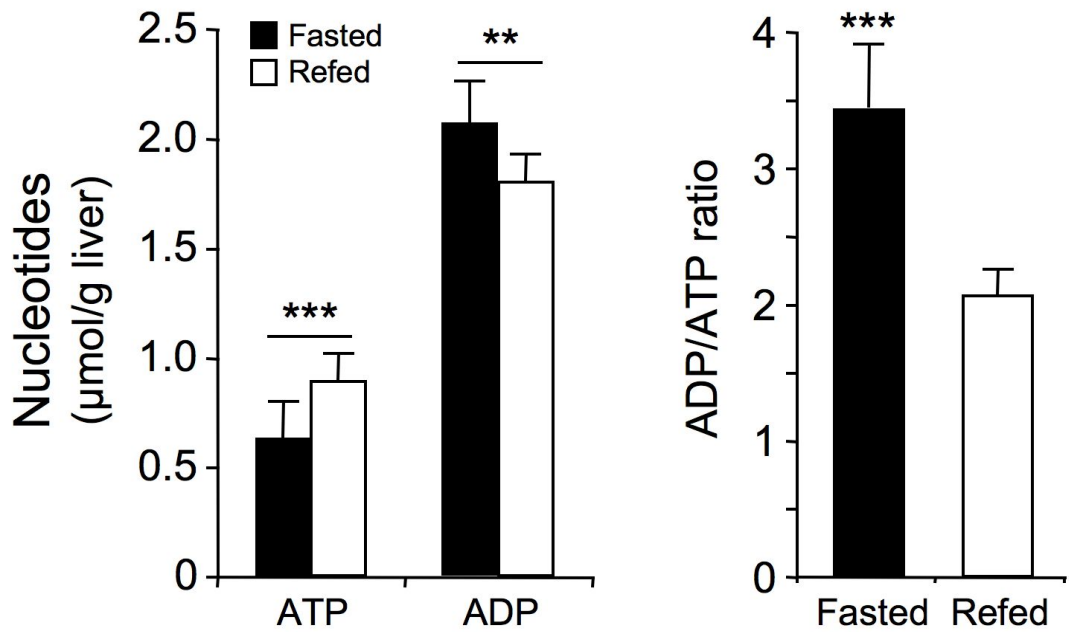

**Figure 2. Effect of the fasting/refeeding transition on the energy state in the liver.** Ten-week-old C57BL6J mice were either fasted for 24 h (Fasted) or fasted for 24 h and then refed a high-carbohydrate diet (Refed) for 3 h. After nutritional manipulation, the livers were quickly collected to determine the ATP and ADP content and ADP/ATP ratios. Data are presented as the means ± SD. N = 10 per group. $^{**}P < 0.01$, $^{***}P < 0.001$ compared to refed mice.

**Figure 3**

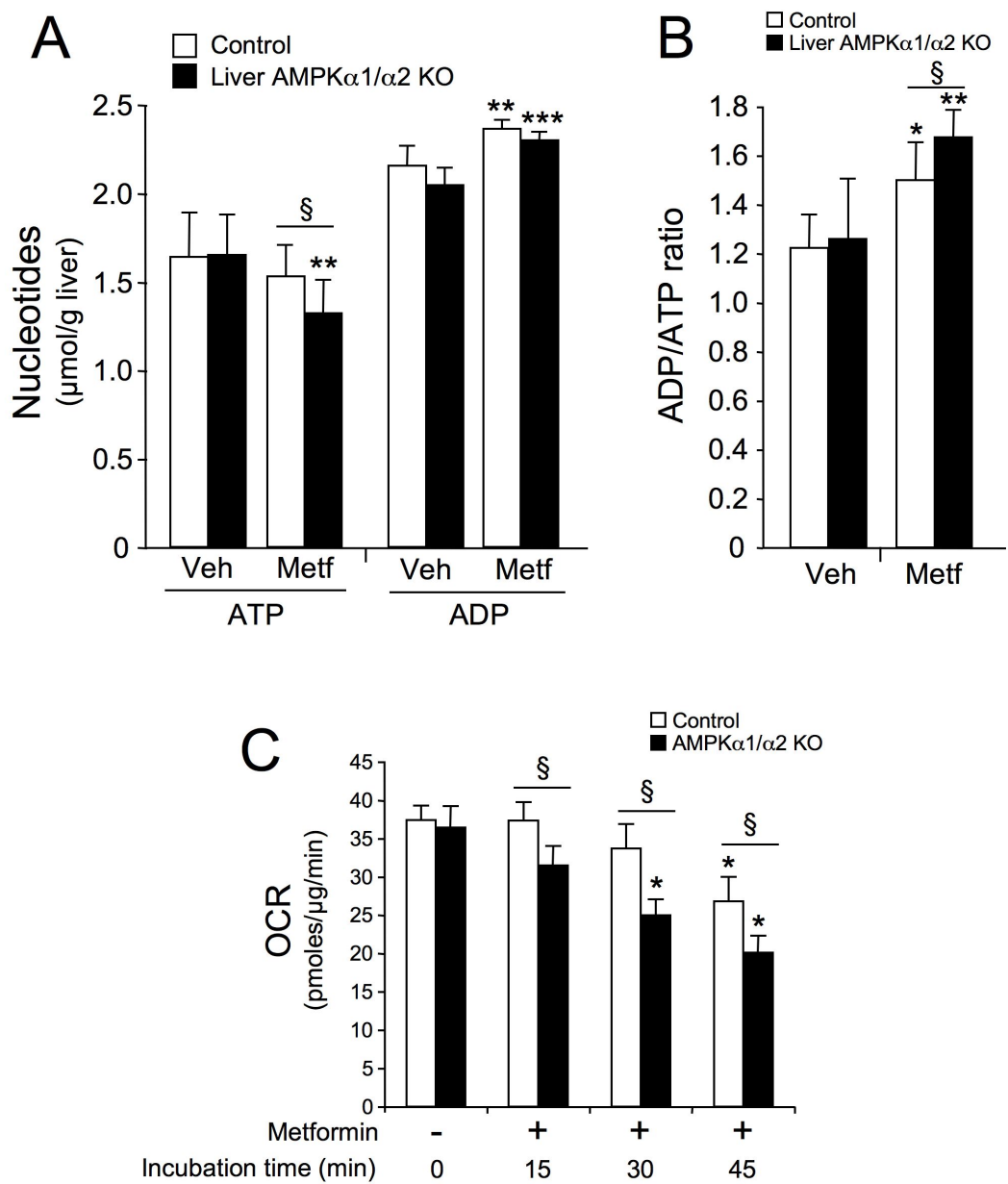

**Figure 3. Liver AMPK-deficient mice are more sensitive to hepatic energy stress.** Ten-week-old control and liver AMPKα1/α2 KO mice (n = 7-8 per group) in the fed state were injected intraperitonaly with saline (Veh) or 200 mg/kg metformin (Metf) to induce hepatic energy stress. After 1 h, livers were quickly collected as described in Experimental Procedures for hepatic ATP and ADP determination. (**A**) Liver ATP and ADP content and (**B**) ADP/ATP ratios are shown for each condition. Data are presented as the means ± SD. $^*P < 0.05$, $^{**}P < 0.01$, $^{***}P < 0.001$ compared to vehicle-treated control or liver AMPKα1/α2 KO mice; $^§P < 0.05$ compared to metformin-treated control mice. (**C**) Effect of metformin on respiration in control and AMPKα1/α2 KO hepatocytes. Control and AMPKα1/α2-deficient mouse primary hepatocytes plated in specialized microplates were switched to glucose-free medium supplemented with lactate and pyruvate (10:1 mM) and 100 nM dexamethasone 1 h prior to measuring respiration. The oxygen consumption rate (OCR, mitochondrial respiration) was monitored using the Seahorse Bioscience XF96 Extra Cellular Flux Analyzer in real time. The OCR was acquired under basal conditions and 15, 30, and 45 min after injection with 1 mM metformin. Results were normalized to total protein/well after completion of the assay. Results are representative of three independent experiments. Data are presented as the means ± SD. $^*P < 0.05$ compared to basal conditions of control or AMPKα1/α2 KO hepatocytes; $^§P < 0.05$ compared to control hepatocytes incubated under the same conditions.

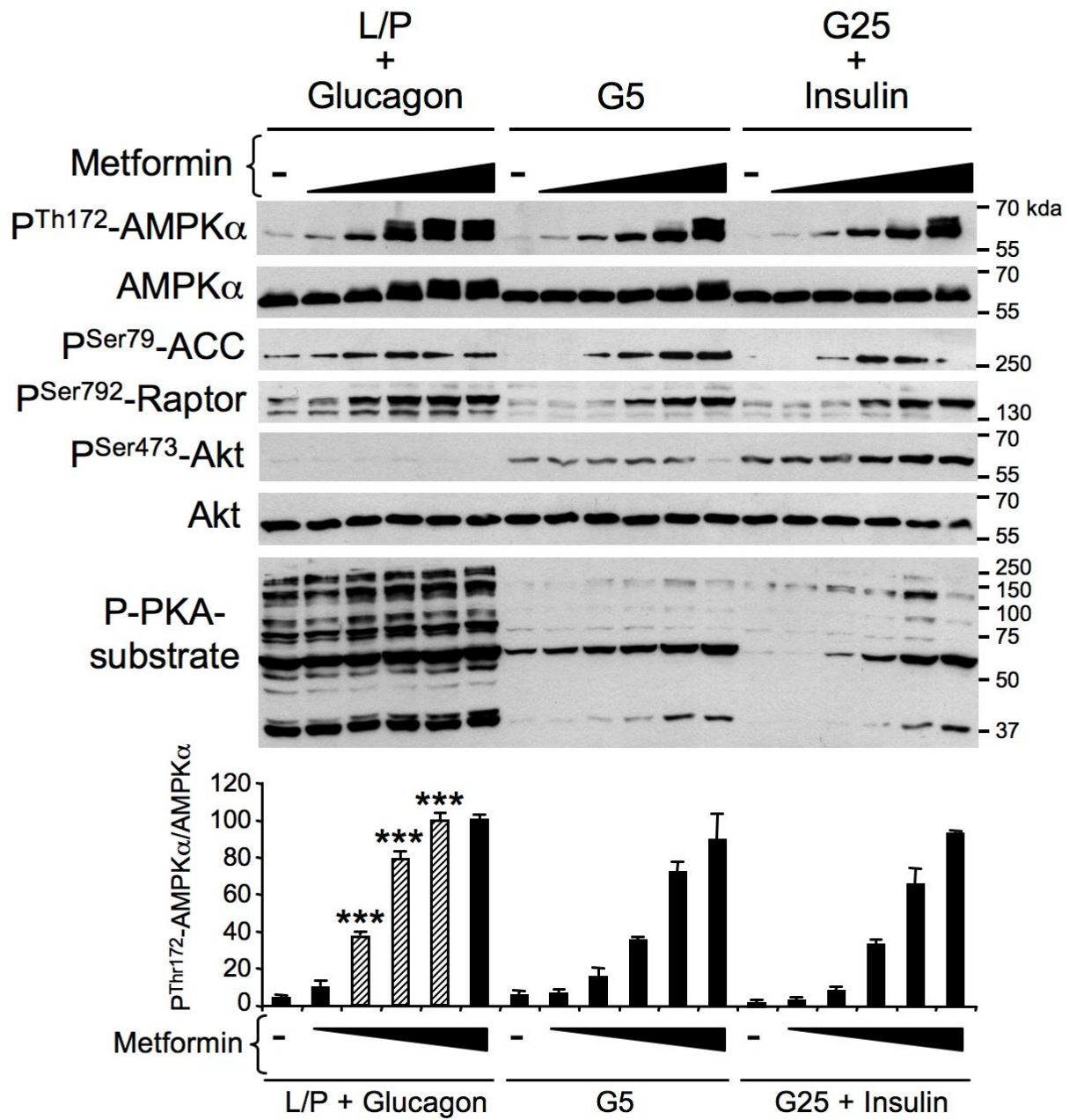

**Figure 4. Effect of fasting-like and refeeding-like culture conditions on energy stress-induced AMPK activation in primary hepatocytes.** Mouse primary hepatocytes were treated with various concentrations of metformin (0, 0.125, 0.25, 0.5, 1, or 2 mM) in glucose-free medium containing 10 mM lactate and 1 mM pyruvate plus 10 nM glucagon (L/P+Glucagon), which mimics the fasting state, in basal medium containing 5 mM glucose alone (G5), or in refeeding-like medium containing 25 mM glucose plus 100 nM insulin (G25+insulin). After 8 h, cells were harvested for western-blot analysis. Immunoblots from hepatocyte lysates were performed using the indicated antibodies. The lower panel represents the P-Thr172-AMPKα/AMPKα ratio from the quantification of immunoblot images. Results are representative of three independent experiments. Data are presented as the means ± SD. ***$P < 0.001$ compared to G5 or G25+Insulin conditions. Hatched bars indicate an increase in the P-Thr172-AMPKα/AMPKα ratio in hepatocytes incubated in L/P+Glucagon relative to those incubated in G5 or G25+Insulin.

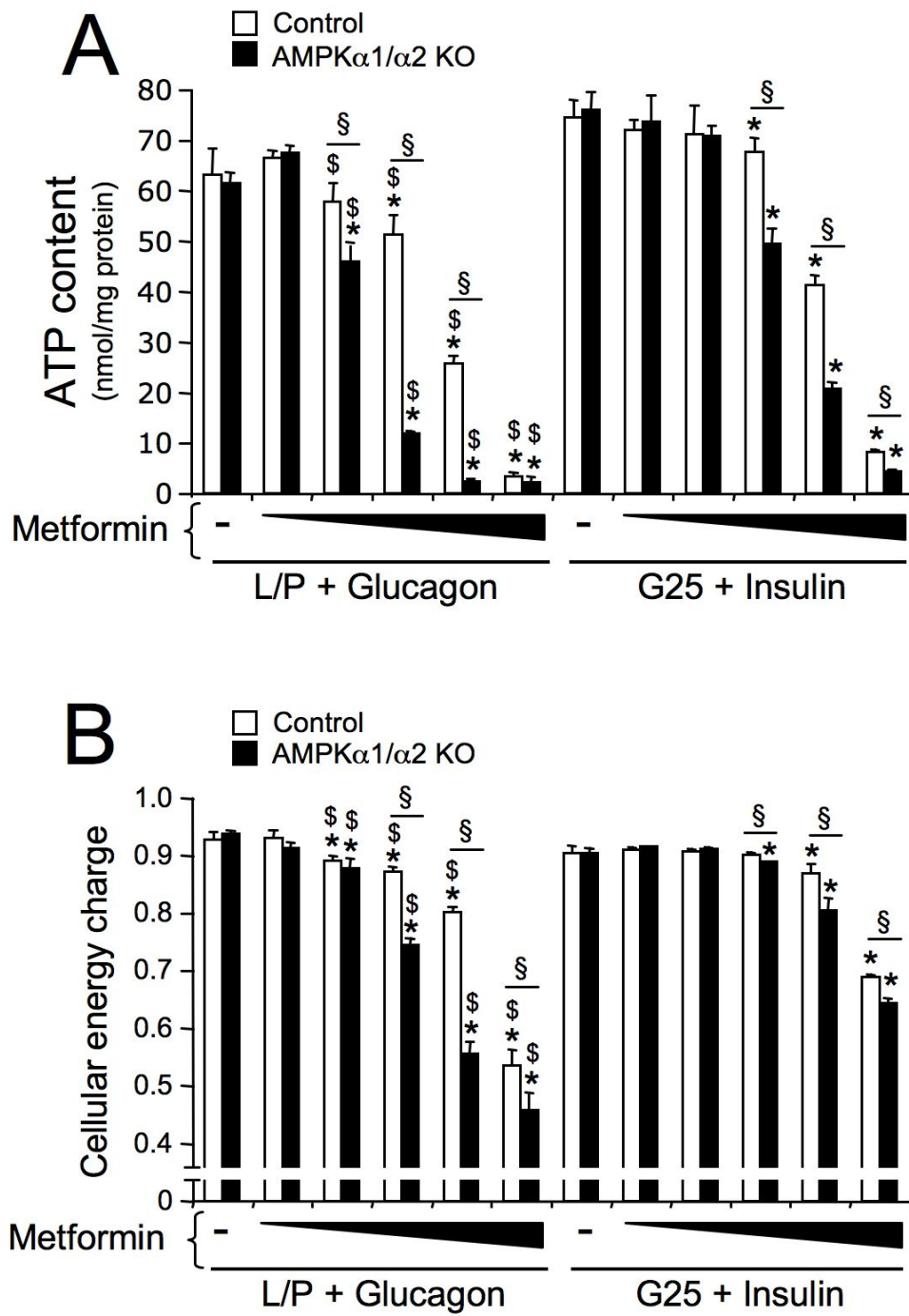

Figure 5. Effect of fasting-like and refeeding-like culture conditions on energy depletion induced by a metformin-induced energy stress in control and AMPK-deficient hepatocytes. Control and AMPKα1/α2 KO mouse primary hepatocytes were treated with various concentrations of metformin (0, 0.125, 0.25, 0.5, 1, or 2 mM) in glucose-free medium containing 10 mM lactate and 1 mM pyruvate plus 10 nM glucagon (L/P+Glucagon), which mimics the fasting state, in basal medium containing 5 mM glucose alone (G5), or in refeeding-like medium containing 25 mM glucose plus 100 nM insulin (G25+insulin). After 8 h, cells were harvested for measurement of the adenine nucleotide content. (**A**) Intracellular ATP content. (**B**) Adenylate energy charge in control and AMPKα1/α2 KO hepatocytes. Results are representative of three independent experiments. Data are presented as the means ± SD. *$P < 0.05$ compared to control or AMPKα1/α2 KO hepatocytes incubated without metformin; §$P < 0.05$ compared to control hepatocytes incubated under the same conditions; $$P < 0.05$ compared to AMPKα1/α2 KO hepatocytes incubated with the same metformin concentration in medium containing G25+Insulin.

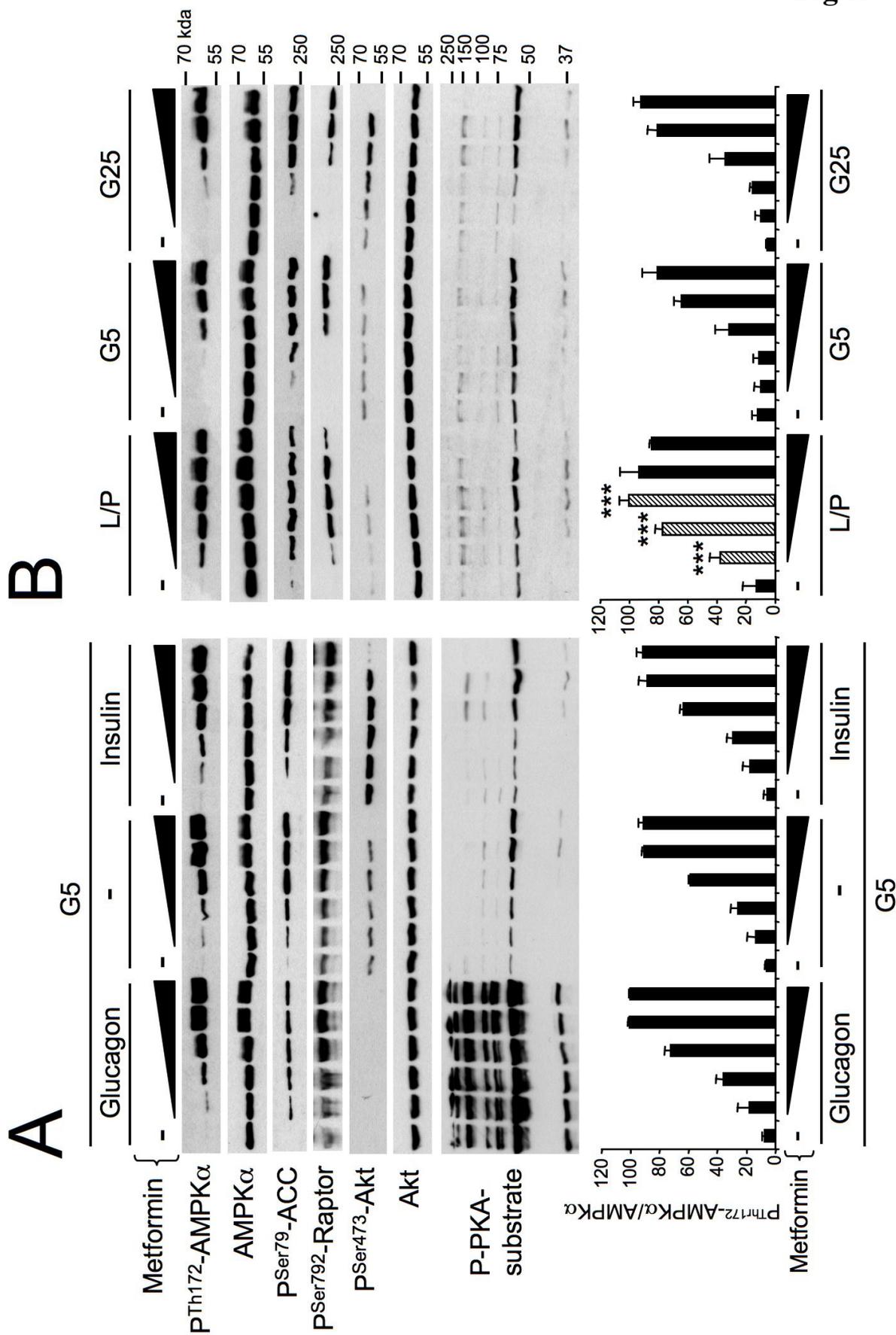

Figure 6

**Figure 6. Effect of glucose and pancreatic hormone levels on energy stress-induced AMPK activation in primary hepatocytes.** Mouse primary hepatocytes were incubated with various concentrations of metformin (0, 0.125, 0.25, 0.5, 1, or 2 mM) and under various culture conditions, including (**A**) 5 mM glucose plus 10 nM glucagon (G5+Glucagon), 5 mM glucose alone (G5), 5 mM glucose plus 100 nM insulin (G5+Insulin), (**B**) glucose-free medium containing 10 mM lactate and 1 mM pyruvate (L/P), 5 mM glucose alone (G5), or 25 mM glucose alone (G25). After 8 h, cells were harvested for western-blot analysis. Immunoblots from hepatocyte lysates were performed using the indicated antibodies. Lower panels represent the P-Thr172-AMPKα/AMPKα ratio from the quantification of immunoblot images. Results are representative of three independent experiments. Data are presented as the means ± SD. $^{***}P < 0.001$ compared to G5 or G25 conditions. Hatched bars indicate an increase in the P-Thr172-AMPKα/AMPKα ratio in hepatocytes incubated in L/P relative to those incubated in G5 or G25.

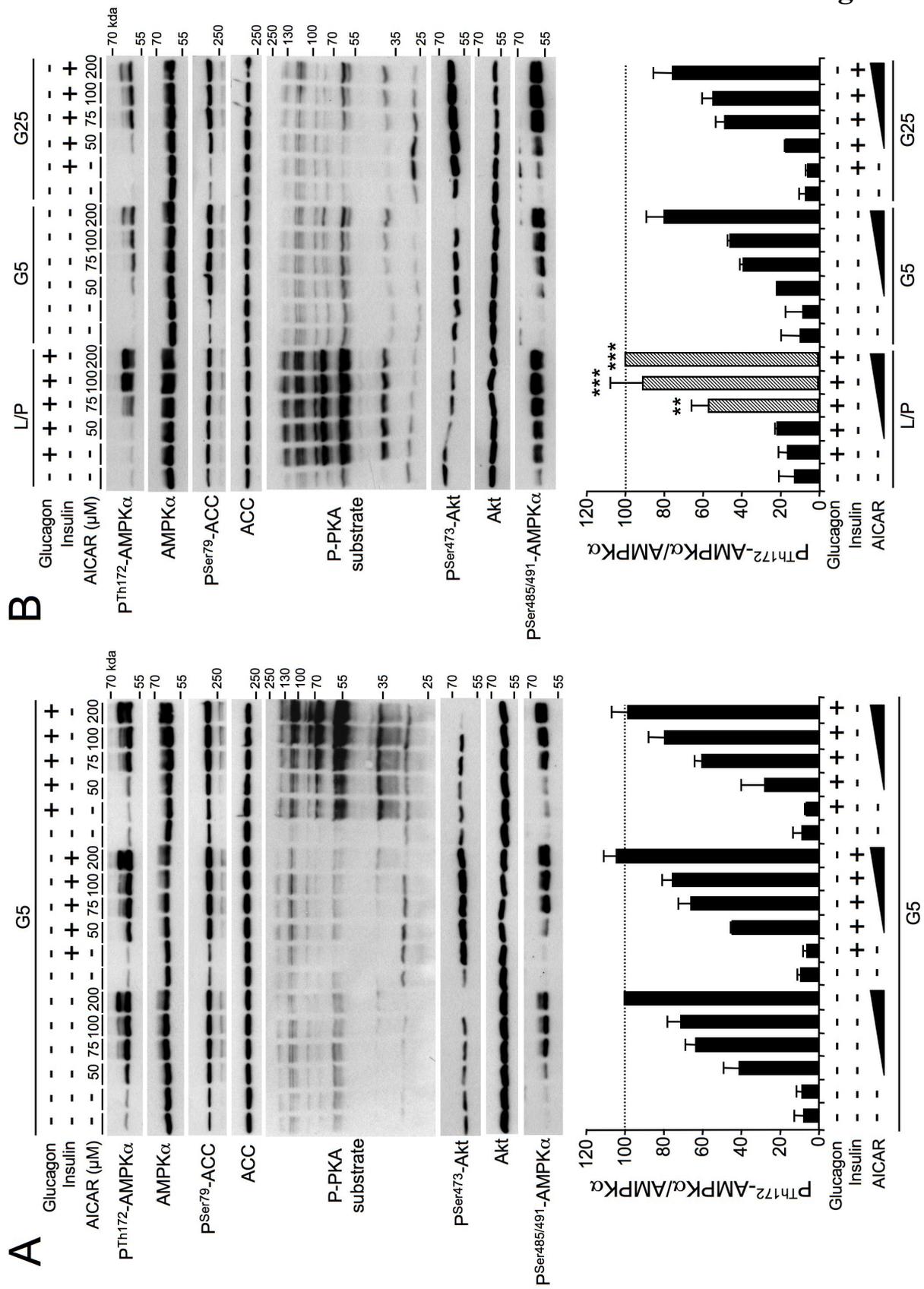

**Figure 7**

**Figure 7. Effect of glucose and pancreatic hormone levels on AICAR-mediated AMPK activation in primary hepatocytes.** Mouse primary hepatocytes were preincubated for 1 h under various culture conditions, including (**A**) 5 mM glucose (G5), 5 mM glucose plus 100 nM insulin, 5 mM glucose plus 100 nM glucagon, (**B**) 10 mM lactate and 1 mM pyruvate without glucose (L/P) plus 10 nM glucagon, 5 mM glucose (G5), or 25 mM glucose (G25) plus 100 nM insulin and then various concentrations of AICAR (0, 50, 75, 100, or 200 μM) were added to the medium. After 8 h, cells were harvested for western-blot analysis. Immunoblots from hepatocyte lysates were performed using the indicated antibodies. Lower panels represent the P-Thr172-AMPKα/AMPKα ratio from the quantification of immunoblot images. Results are representative of three independent experiments. Data are presented as the means ± SD. **$P < 0.01$, ***$P < 0.001$ compared to G5 or G25 conditions. Hatched bars indicate an increase in the P-Thr172-AMPKα/AMPKα ratio in hepatocytes incubated in L/P relative to those incubated in G5 or G25.

# Figure 8

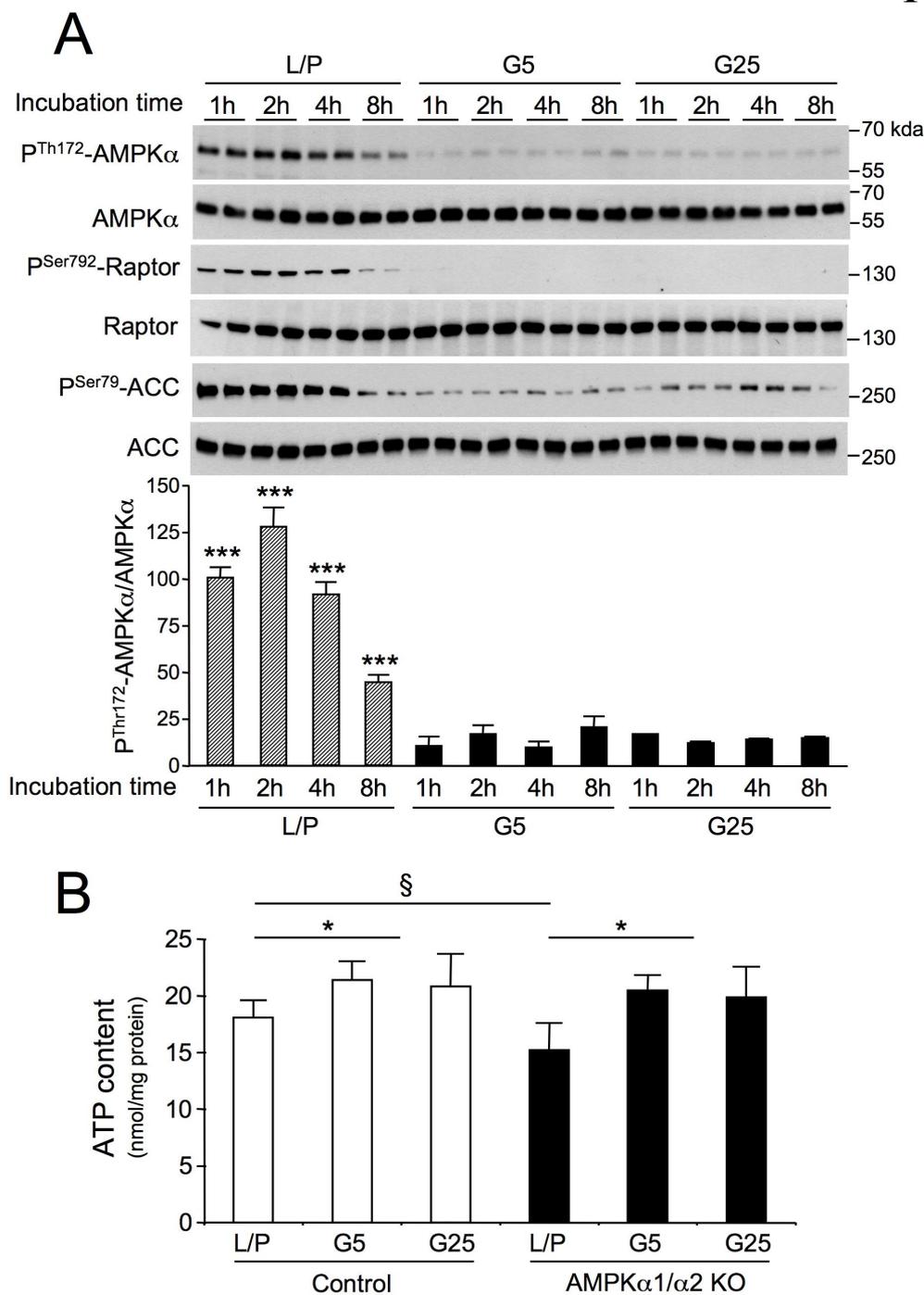

**Figure 8. Time course of the effect of changing glucose levels on AMPK activity in primary hepatocytes.** (**A**) Mouse primary hepatocytes were incubated under various culture conditions, including glucose-free medium containing 10 mM lactate and 1 mM pyruvate (L/P), 5 mM glucose (G5), or 25 mM glucose (G25). After 1, 2, 4 or 8 h, cells were harvested for western-blot analysis. Immunoblots from hepatocyte lysates were performed using the indicated antibodies. The lower panel represents P-Thr172-AMPKα/AMPKα ratios from the quantification of immunoblot images. Results are representative of three independent experiments. Data are presented as the means ± SD. $^{***}P < 0.001$ compared to G5 or G25. Hatched bars indicate an increase in the P-Thr172-AMPKα/AMPKα ratio in hepatocytes incubated in L/P relative to those incubated in G5 or G25. (**B**) Control and AMPKα1/α2 KO primary hepatocytes were incubated in glucose-free medium containing L/P or G5 or G25. After 2 h, cells were harvested for the measurement of intracellular ATP content. Data are presented as the means ± SD. N = 6. $^{*}P < 0.05$, L/P compared to G5 or G25. $^{§}P < 0.05$, Control compared to AMPKα1/α2 KO hepatocytes incubated in L/P.
22

Figure 9

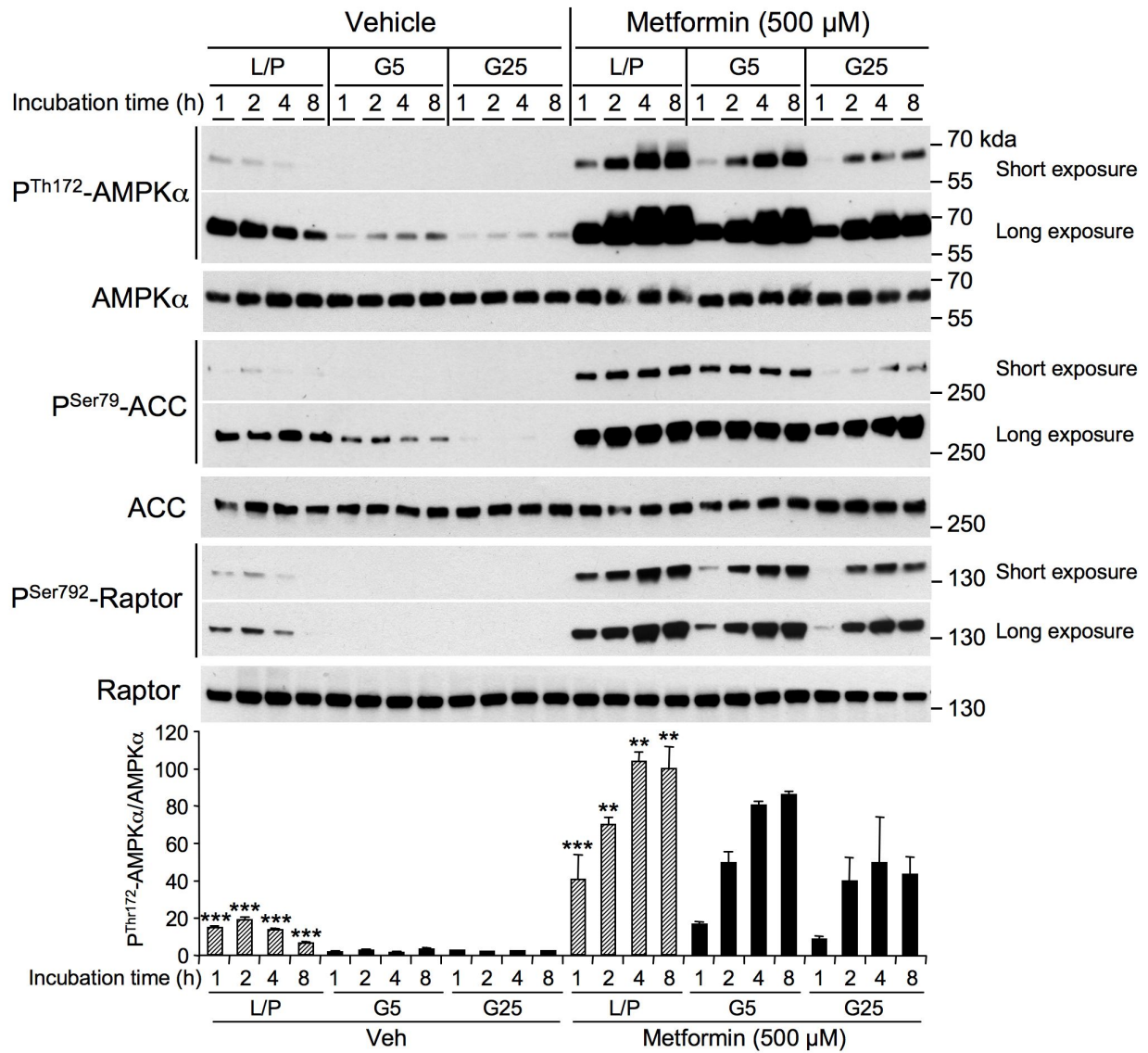

**Figure 9. Effect of glucose levels on the kinetics of metformin-mediated AMPK activation in primary hepatocytes.** (A) Mouse primary hepatocytes were incubated without (Vehicle) or with 500 μM metformin under various culture conditions, including glucose-free medium containing 10 mM lactate and 1 mM pyruvate (L/P), 5 mM glucose (G5), or 25 mM glucose (G25). After 1, 2, 4, or 8 h, cells were harvested for western-blot analysis. Immunoblots from hepatocyte lysates were performed using the indicated antibodies. The lower panel represents the P-Thr172-AMPKα/AMPKα ratio from the quantification of immunoblot images. Data are presented as the means ± SD. n = 3. $^{**}P < 0.01$, $^{***}P < 0.001$ compared to G5 or G25. Hatched bars indicate an increase in the P-Thr172-AMPKα/AMPKα ratio in hepatocytes incubated in L/P relative to those incubated in G5 or G25.

# Supplemental Figures

**Glucose availability but not changes in pancreatic hormones sensitizes hepatic AMPK activity during nutritional transition in rodents**

Camille Huet, Nadia Boudaba, Bruno Guigas, Benoit Viollet and Marc Foretz

# Figure S1

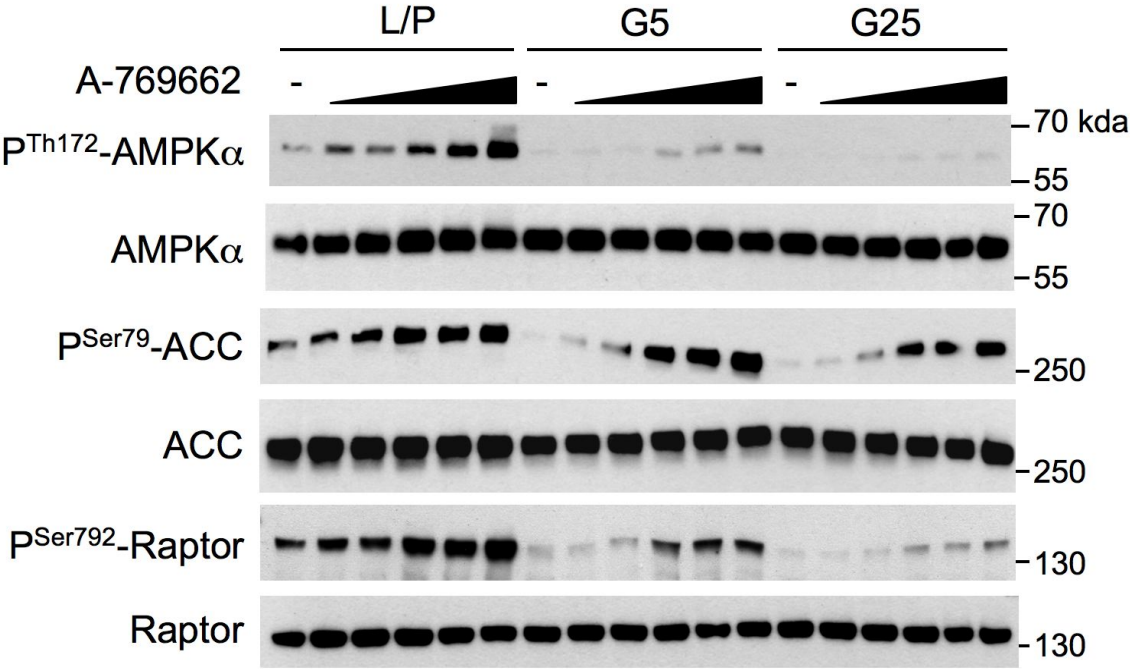

**Figure S1. Effect of glucose levels on A-769662-mediated AMPK activation in primary hepatocytes.** Mouse primary hepatocytes were incubated with various concentrations of A-762669 (0, 0.1, 0.3, 1, 3, or 10 µM) under various culture conditions, including glucose-free medium containing 10 mM lactate/1 mM pyruvate (L/P), 5 mM glucose (G5), or 25 mM glucose (G25). After 8 h, cells were harvested for western-blot analysis. Immunoblots from hepatocyte lysates were performed using the indicated antibodies. Presented blots are representative of two independent experiments.

# Figure S2

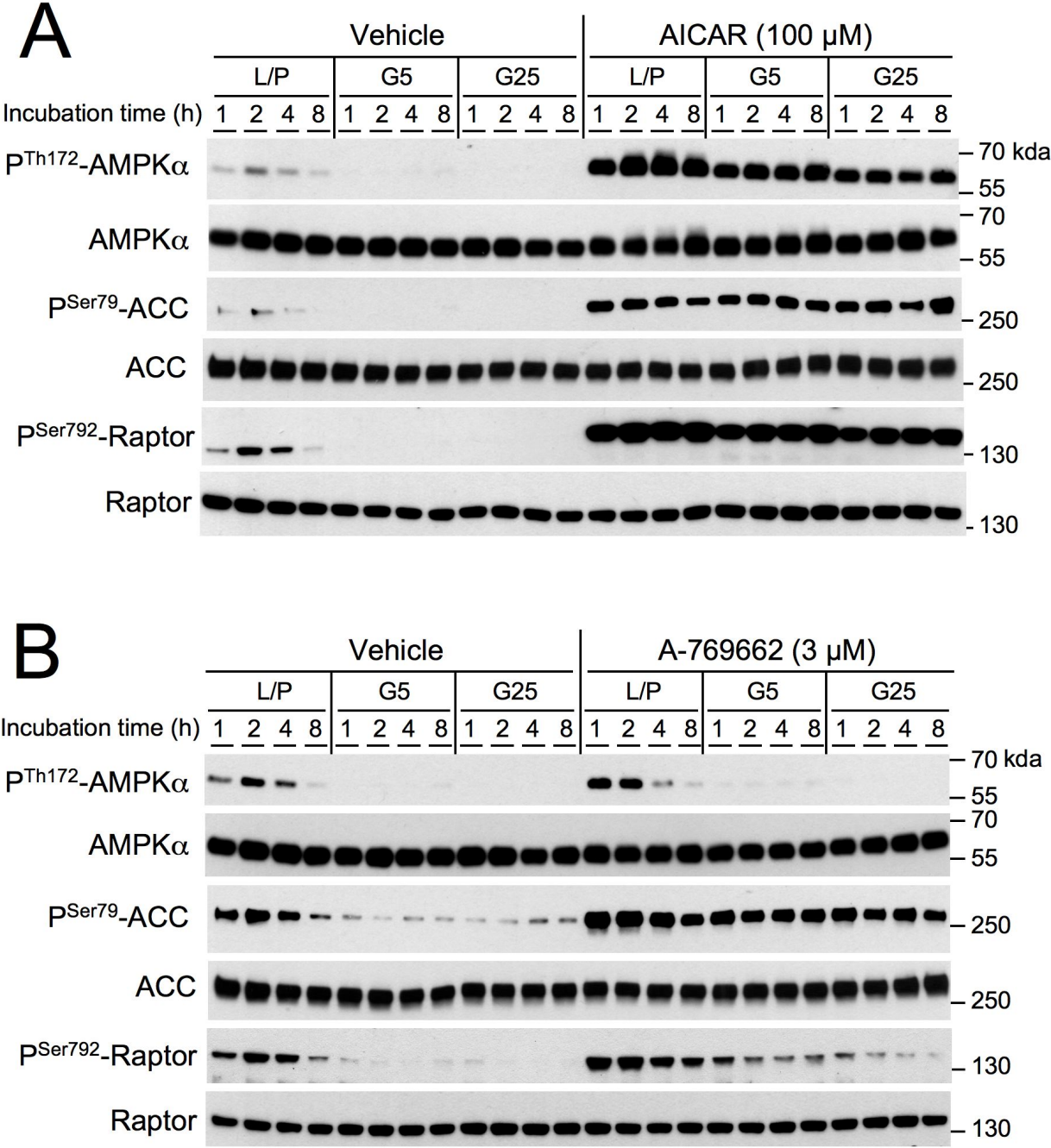

**Figure S2. Effect of glucose levels on the kinetics of AICAR- and A-769662-mediated AMPK activation in primary hepatocytes.** Mouse primary hepatocytes were incubated without (Vehicle) or with 100 μM AICAR (**A**) or 3 μM A-769662 (**B**) under various culture conditions, including glucose-free medium containing 10 mM lactate/1 mM pyruvate (L/P), 5 mM glucose (G5), or 25 mM glucose (G25). After 1, 2, 4, or 8 h, cells were harvested for western-blot analysis. Immunoblots from hepatocyte lysates were performed using the indicated antibodies. Presented blots are representative of two independent experiments.